\documentclass[usenatbib]{mn2e}
\usepackage{graphicx}
\usepackage{stfloats}
\usepackage{amsmath}
\usepackage{amssymb}
\title[Gravitational Wave Heating of Stars and Accretion Disks]{Gravitational Wave Heating of Stars and Accretion Disks}

\author[Li, Kocsis and Loeb]{Gongjie Li, Bence Kocsis, Abraham Loeb \\ Institute for Theory \& Computation, Harvard-Smithsonian Center for Astrophysics, Cambridge, MA, USA}

\begin{document}
\topmargin-0.5cm
\bibliographystyle{mn2e}
\maketitle
%\date{Accepted 1988 December 15. Received 1988 December 14; in original form 1988 October 11}
%\pagerange{\pageref{firstpage}--\pageref{lastpage}} \pubyear{2002}

\newcommand{\apj}{ApJ}
\newcommand{\apjl}{ApJL}
\newcommand{\apjs}{ApJS}
\newcommand{\mnras}{MNRAS}
\newcommand{\aap}{AAP}
\newcommand{\prd}{PRD}
\newcommand{\aj}{AJ}
\newcommand{\pasp}{PASP}
\newcommand{\araa}{ARA\&A}
\newcommand{\nat}{Nature}
\newcommand{\be}{\begin{equation}}
\newcommand{\ee}{\end{equation}}
\newcommand{\bea}{\begin{eqnarray}}
\newcommand{\eea}{\end{eqnarray}}

\def\Mpc{\rm Mpc}
\def\Mbh{M_{\rm BH}}

\def\Msun{M_{\rm \odot}}
\def\kpc{\rm kpc}
\newcommand{\comment}[1]{}

%%%%%%%%%%%%%%%%%%%%%%%%%%%%%%%%%%%%%%%%%%

%---------------------------------------------------------------

\begin{abstract}
We investigate the electromagnetic (EM) counterpart of gravitational waves (GWs) emitted by a supermassive black hole binary (SMBHB) through the viscous dissipation of the GW energy in an accretion disk and stars surrounding the SMBHB. We account for the suppression of the heating rate if the forcing period is shorter than the turnover time of the largest turbulent eddies. We find that the viscous heating luminosity in $0.1\Msun$ stars can be significantly higher than their intrinsic luminosity, but still too low to be detected for extragalactic sources. The relative brightening is small for accretion disks.
\end{abstract}


%-----------------------------------------------------------------

\section{Introduction}
\label{s:intro}
The coalescence of supermassive black hole binaries (SMBHBs) generates gravitational waves (GW) which are a primary source for the proposed Laser Interferometric Space Antenna (LISA\footnote{http://lisa.nasa.gov/}). SMBHBs are inevitable outcomes of galaxy mergers. Spatially-resolved active galactic nuclei have been observed \citep{Konossa03, Bianchi08, Green10, Koss11, Fabbiano11}. In addition, spectroscopic surveys \citep{Comerford09, Smith10, Liu10} and observations that combine ground-based imaging show numerous systems containing compelling SMBHB candidates with pc to kpc separations \citep{Rodriguez06, Liu102, Shen11, Fu11, McGurk11}. Hydrodynamic simulations of galaxy mergers also predict SMBHB pair formation \citep[e.g.][]{Escala04, Escala05, DiMatteo05, Robertson06, Hopkins06, Callegari09, Colpi09, Blecha12}.

Electromagnetic (EM) counterparts to GW sources complements the GW detection by determining the host galaxy redshift and the environment of the sources \citep{Kocsis06, Phinney09}. A large variety of EM signatures have been proposed to accompany the coalescence of SMBHBs \citep{Schnittman11, Haiman09}. In the pre-merger phase, the torques of the SMBHB excavates a hollow region in the disk and leads to periodic accretion across the gap on the orbital timescale \citep{Cuadra09, MacFadyen08, Hayasaki08}. After the merger, the recoil of the black hole remnant and its sudden mass loss due to the final GW burst produce shocks in the accretion disk which lead to EM signals \citep{Bode07, Lippai08, Schnittman08, Shields08, ONeill09, Rossi10}. The recoil of the black hole remnant changes the tidal disruption rate of stars due to the refilling of the loss cone and the wandering of black hole remnant \citep{Stone111, Stone11, Li12}. Finally, the infall of gas onto the black hole remnant produces an EM afterglow \citep{Milosavljevi05, Tanaka12}.

In this paper, we consider the viscous dissipation of GWs generated by a SMBHB in a neighboring gaseous medium. In particular, the velocity shear induced by GWs in the gas is damped by viscosity. The dissipated GW energy turns into heat, and produces an electromagnetic flare. Unlike other EM counterparts, the brightening here follows promptly within a few hours to days after the coalescence of the SMBHB \citep{Kocsis08}. The effect provides a unique test of general relativity for the interaction of GWs with matter. In \S~\ref{s:met} and \ref{s:res} we investigate GW dissipation in a gaseous accretion disk and stars in the vicinity of the SMBHB. We examine the suppression of the effect if the forcing period is shorter than the turnover time of the largest eddies \citep{Krolik10}, in analogy to a similar treatment of tidal heating in binary stars \citep{Zahn66, Goldreich77}. Finally, we discuss our conclusions and their implications in \S~\ref{s:conc}.

\section{Method}
\label{s:met}
We start by presenting our approach for estimating the GW heating inside an accretion disk and stars due to turbulent viscosity. Following \citet{Kocsis08}, we approximate the GW luminosity by matching the Newtonian inspiral luminosity prior to merger ($t<0$), the peak luminosity at the merger ($t = 0$) and the decay luminosity afterwards ($t>t_1$), where $t_1$ can be fixed from this matching procedure. Specifically, in the Newtonian inspiral regime, the luminosity is
\be
L_{\rm GW\,inspiral} = \frac{32}{5}\frac{G^4}{c^5}\frac{M^3\mu^2}{a^5}, \\
\ee
where $M=M_1+M_2$ is the sum of the masses of the SMBHB members, $\mu = M_1M_2/M$ is the reduced mass of the SMBHB and $a$ is the separation between the SMBHB, which can be expressed as
\be
a = \left[ \frac{256}{5} \frac{G^3}{c^5} \mu M^2 (t_1-t) \right]^{1/4},\\
\ee
assuming a circular orbit. The peak luminosity is approximated from numerical simulations \citep{Berti07, Buonanno07} as
\be
L_{\rm GW\,peak} \approx 10^{-3}\frac{c^5}{G}\Big(\frac{\mu}{M}\Big)^2,\\
\ee
and the ringdown luminosity is set to be
\be
L_{\rm GW\,ringdown} = L_{\rm GW\,peak} \exp\hspace{-2pt}\left(-\frac{c (t-t_1) }{5 R_{\rm g}}\right),\\
\ee
where $R_{\rm g} = G M / c^2$ is the gravitational radius of the SMBHB. The peak luminosity is modified by a factor of two \citep{Berti07, Buonanno07} due to different magnitudes and orientation of the spin of the SMBHB. In this paper, we assume the masses of the two black holes are the same.

With the approximated expression of GW luminosity as a function of time, the dissipation of GW energy inside a viscous medium can be calculated by solving the weak-field Einstein equation \citep{Hawking66, Weinberg72}:
\be
\dot{e}_{\rm heat} = \frac{16 \pi G \eta}{c^2}e_{\rm GW},\\
\label{e:eheat}
\ee
where $\dot{e}_{\rm heat}$ is the dissipation rate, $\eta$ is the dynamical viscosity and $e_{\rm GW}$ is the GW energy density. $e_{\rm GW}$ can be obtained from $e_{\rm GW}=Y(\theta)\frac{L_{\rm GW}}{4\pi c r^2}$, where $\theta$ is the angle relative to the total angular momentum vector, $Y(\theta) = 5/2 [\sin^8(\theta/2) + \cos^8 (\theta/2) ]$. We use the average value $\langle Y\rangle=1$ below. With $L_{\rm GW}$ derived, the only unknown parameter is the dynamical viscosity of the medium that the GW passes through. The dissipation rate of the GW energy gives the heating rate of any gaseous medium such as an accretion disk and stars.

Next, we estimate the dynamical viscosity for stars. We use stellar models produced by Modules for Experiments in Stellar Astrophysics (MESA\footnote{http://mesa.sourceforge.net/}) \citep{Paxton11}, a 1D stellar evolution code, and we consider stellar models, whose properties are included in Table \ref{tab: t1}. We associate the dynamical viscosity with the mixing length theory diffusion coefficient, which is directly provided in the simulated models by MESA. When the period of the driving force is smaller than the largest eddy turnover time, the eddy viscosity depends on the ratio of the period to the largest eddy turnover time in one of two possible ways:
\be \label{e:eta1}
\eta = \eta_i~{\rm min}~\Big[\Big( \frac{\tau_{\rm GW}}{2 \tau_l} \Big), 1 \Big] ,\\
\ee
or
\be \label{e:eta2}
\eta = \eta_i~{\rm min}~\Big[\Big( \frac{\tau_{\rm GW}}{2 \pi \tau_l} \Big)^2, 1 \Big] ,\\
\ee
where $\eta_i$ is the intrinsic viscosity in the absence of shear force with short period, $\tau_l$ is the largest eddy turnover timescale and $\tau_{\rm GW}$ is the shear force period, which is calculated as $2\pi/\omega_{\rm GW}$, where $\omega_{\rm GW}=2\sqrt{GM/a^3}$ in the inspiral phase $a<6R_{\rm g}$ and $0.25/(G M / c^3)]$ after the ringdown, and extrapolate linearly during the transition according to \citet{Buonanno07}. The viscosity scaling given by Eq.~(\ref{e:eta1}) is discussed in \citet{Zahn66, Zahn891, Zahn892} and Eq.~(\ref{e:eta2}) in \citet{Goldreich77, Goldreich89}. Observations are more consistent with Zahn's scaling for pulsating stars in the red edge of the instability strip \citep{Gonczi82}, for tidal circularization of binary stars \citep{Verbunt95, Meibom05}, while the damping of the solar p-mode oscillations is more consistent with the Goldreich's scaling \citep{Goldreich88, Goldreich94}. Recently, \citet{Penev09} studied turbulent viscosity in low mass stars using the perturbative approach of \citet{Goodman97}, taking into account compressible fluid and anisotropic viscosity. Their simulation suggests a linear scaling. However, \citet{Ogilvie12} found results more consistent with Goldreich's scaling when studying the limit of a low amplitude short oscillation period shear. We considered both scalings for stars in this paper.

With the viscosity for stars and $L_{\rm GW} (t)$ in hand, the GW heating rate can be estimated using Eq.~(\ref{e:eheat}). The EM luminosity increase can be estimated by solving the radiative transfer equation:
\bea
t_c(r) \frac{d}{dt}\Delta f(r)+\Delta f(r) = \dot{e}_{\rm heat},
\label{e:fstar1}\\
L_{\rm GWH} = \int_{\rm star}~\Delta f(r)~dV,
\label{e:fstar2}
\eea
where $\Delta f (r)$ is the excess EM signal produced per unit volume as a function of location in the star, $L_{\rm GWH}$ is the excess EM luminosity associated to GW heating, and $t_c (r)$ is the cooling time as a function of the location, which characterizes the time it takes for heat to travel to the surface. We estimate the latter by taking the integral of the minimum of the photon diffusion time, $ dr/c \times [\tau(r)-(R_*-r)\frac{d\tau(r)}{dr}]$, and the turbulent convection time, $dr/v_c(r)$, in each spherical shell inside the star, where the optical depth, $\tau(r)$, and the convective velocity, $v_c(r)$ are obtained from the MESA simulation, and $R_*$ is the radius of the star.

Finally, we estimate the heating in accretion disks. We adopt the geometrically thin, optically thick, standard accretion disk model, where the angular momentum transport is associated with the internal stresses due to turbulence \citep{Shakura73, Novikov73}. Heat is dissipated locally by turbulent viscosity, and transported vertically outward by photon diffusion or advection. Specifically, the viscosity of the accretion disk is
\be
\eta_i (r) = \frac{2}{3} \frac{\alpha P(r)}{\Omega(r)}, \\
\ee
where $\Omega^2(r) = GM/r^3$ is the angular velocity, $\alpha$ is a constant which we assume to be 0.3 \citep{King07}, and $P$ is the total (gas+radiation) pressure in the $\alpha$ disk model, and gas pressure in the $\beta$ model. In these models, the physical characteristics of the disk is fixed by the following parameters: the accretion rate in Eddington units ($\dot{m}$), the radiation efficiency ($\epsilon$), and the SMBHB mass (M) \citep{Goodman03, Goodman04}. We set $\dot{m}$ to be 0.1, $\epsilon$ to be 0.1, and discuss the effects caused by different SMBHB masses.

Similarly to stars, we account for the frequency dependence of viscosity when the period of the driving force is smaller than the largest eddy turnover time, and estimate the effective viscosity according to the perturbative methods as discussed in \citet{Goodman97}.  Specifically, for incompressible fluid with isotropic viscosity, the viscosity as a function of driving force frequency can be expressed in terms of the frequency spectrum of the average kinetic energy per unit mass. For accretion disks, where the Kolmogorov scalings may not be applicable to obtain the energy spectrum, we adopt the energy spectrum from recent magnetohydrodynamic (MHD) disk simulations. \citet{Flock11} present a full $2\pi$ three dimensional simulation on a stratified accretion disk, where the turbulence is driven by magnetorotational instability (MRI) and the kinetic spectra is obtained in the $\phi$ direction, and \citet{Fronmang10} investigate the MRI in a shearing box with zero net flux. \citet{Flock11} and \citet{Fronmang10} estimate the kinetic energy spectrum exponent to be $11/9$ and $1.5$, respectively. We estimate the viscous heating in accretion disk with the \citet{Flock11} exponent as well as the Kolmogorov scaling exponent $2$.

Similarly to the calculation for stars, the GW heating rate can be estimated using Eq.~(\ref{e:eheat}). The corresponding EM signals can be estimated by solving the radiative transfer equation following \citet{Kocsis08}:
\bea
t_c (r) \frac{d}{dt} \Delta F (r,t)+\Delta F(r,t) = H \dot{e}_{\rm heat}(r,t),
\label{e:fdisk1} \\
L_{\rm GWH}(t) = \int_{r_{\min}}^{r_{\max}}~2\pi r \Delta F(r,t) ~dr,
\label{e:fdisk2}
\eea
where $F(r)$ is the excess EM flux due to GW heating in the accretion disk, $H$ is the scaleheight, $L_{\rm GWH}$ is the corresponding excess EM luminosity, and $t_c(r)$ is the cooling time. Here, we assume the disk is face on, and account for the different light-travel time from different annuli in the disk. The brightening can be somewhat larger in an inclined or edge-on configuration (by up to a factor of $\sim 3$) where the peak GW flux is observed coincidentally at the inner and outer radii along the line of sight  \citep{Kocsis08}.

\section{Results}
\label{s:res}
\begin{figure}
\includegraphics[width=3.3in, height=2.7in]{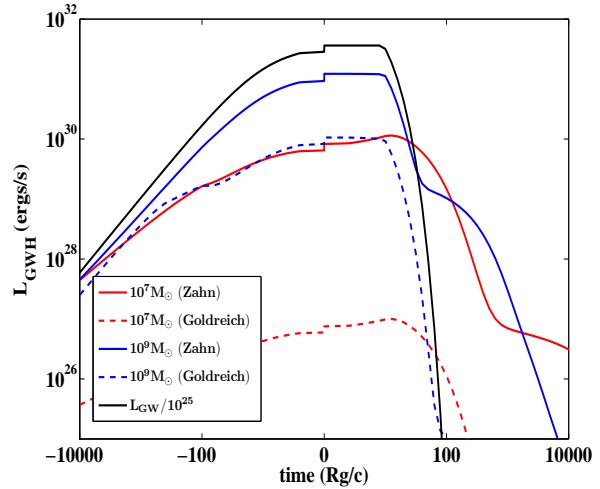}
\caption{\label{e:f1} The light curve of a GW heated star (based on star model 2 with an intrinsic luminosity: $L=2.6\times10^{30} \rm erg~s^{-1}$). The time axis is in units of $R_{\rm g}/c$, and is shown on a logarithmic scale at both negative and positive values (causing the discontinuity at $t=0$). The star is located 5 tidal radii away from the SMBHB ($320$ and $15\,R_{\rm g}$ for a $10^7\, \Msun$ and $10^9\,\Msun$ SMBHB, respectively). The black line indicates the GW luminosity scaled down by 25 order of magnitude in order to fit in this figure. The red and blue lines indicate the light curve of a star surrounding a $10^7$ and a $10^9\,\Msun$ SMBHB, respectively, with solid and dashed lines corresponding to the viscosity dependence with $(\tau_{\rm GW}/2 \tau_l)$ and $(\tau_{\rm GW}/2 \pi \tau_l)^2$, respectively. The light curve closely tracks the GW light curve. Interestingly, the peak luminosity surrounding the $10^9\,\Msun$ SMBHB is much higher than the intrinsic luminosity of this star.}
%\vspace{0.1cm}
\end{figure}

First, we consider the GW heating of nearby stars. As an example, we examine the GW heating light curve for a $0.1~\Msun$ star (stellar model 2) surrounding an $M = 10^7$ or an $10^9 \Msun$ SMBHB, respectively. Using Eqs.~(\ref{e:fstar1}) and (\ref{e:fstar2}), we calculate $F(t)$ and plot the GW heating light curve in Figure~\ref{e:f1}. We assume that the star is located at $d=5$ tidal radii from the SMBHB (corresponds to $320$ and $15\,R_{\rm g}$ for a $10^7$ and a $10^9~\Msun$ SMBHB, respectively). Note that since the GW luminosity is proportional to $(d/R_{\rm g})^{-2}$, the GW heating effect is much larger around more massive SMBHBs because the viscosity suppression for a high mass SMBHB is smaller.

Figure~\ref{e:f1} shows that the excess luminosity of the star surrounding the $10^9 \Msun$ SMBHB is much higher than the intrinsic luminosity of the star ($L=2.6\times10^{30} \rm erg~s^{-1}$). In fact, the net dissipated GW energy can exceed the gravitational binding energy near the stellar surface, and could generate a stellar wind. However, as the viscosity is strongly suppressed in the stellar interior ($\frac{\tau_{\rm GW}}{\tau_l (r) }\ll 1$ for $r\lesssim0.99R_{star}$), the heating effect is negligible to the star as a whole. In addition, these stars are very faint; the absolute peak GW heating luminosity in the star is typically too faint to be observed outside of the Galaxy.

Since the turnover time of turbulent eddies is much longer in the interior of the star than that at the surface, the energy is mostly dissipated at the surface. Since the cooling time near the surface ($\sim 200~{\rm s}$) is short compared to the peak GW timescale ($\sim10\,R_g/c\sim500\,{\rm s}\,M_{\rm BH}/10^7\,\Msun$), the light curve of the star closely tracks the luminosity curve of the GW.
When the GW driving period is shorter than the eddy turnover time, the viscosity caused by the eddy depends on the ratio $\tau_{\rm GW}/\tau_l$, where the exact scaling is uncertain as discussed in \S~\ref{s:met}. For stars surrounding a $10^9\, \Msun$ SMBHB, the differences between the two scalings are smaller as the period of peak GW emission for this SMBHB mass is more comparable to the surface eddy turnover time in a $0.1~\Msun$ star.

\begin{table}
\caption{Properties of stellar models}
\begin{tabular}{|l||l|l|l|l|l|}
\hline
No. 	&	Mass	&	Metallicity	 &	Radius 	&	Luminosity 	&	Age	\\
    & ($M_{\odot}$) & (Z) & ($R_{\odot}$) & ($L_{\odot}$) &  (yrs) \\
\hline
1	&	0.1	&	0.16	&	3.3	&	0.00079	&	$2\times10^4$	\\
2	&	0.1	&	0.16	&	3	&	0.00066	&	$5\times10^6$	\\
3	&	0.1	&	0.16	&	0.57	&	$5.5\times10^{-5}$	&	$2\times10^9$	\\
4	&	0.1	&	0.01	&	2.4	&	0.43	&	$2\times10^4$	\\
5	&	0.1	&	0.01	&	0.44	&	0.022	&	$5\times10^6$	\\
6	&	0.1	&	0.01	&	0.12	&	0.0012	&	$2\times10^9$	\\
7	&	100	&	0.04	&	21	&	$1.4\times10^6$	&	$1\times10^4$	\\
8	&	100	&	0.04	&	36	&	$1.7\times10^6$	&	$1\times10^6$	\\
9	&	100	&	0.04	&	960	&	$2.1\times10^6$	&	$2\times10^6$	\\
\hline
\end{tabular}
\medskip\\
\label{tab: t1}
\end{table}

To examine the influence of the GW heating in different types of stars, we consider stellar models of different stellar masses and ages as included in Table 1. We include the extreme cases with $0.1\,\Msun$ and $100\,\Msun$ stars. We plot the ratio of the peak heating luminosity to the intrinsic luminosity for different stellar models in Figure 2 with Zahn's scaling. We find that the influence of the GW heating is more significant as the metallicity of the star increases, and GW heating is not significant for very massive ($M_*\geq100M_{\odot}$) stars.

\begin{figure}
\includegraphics[width=3.5in]{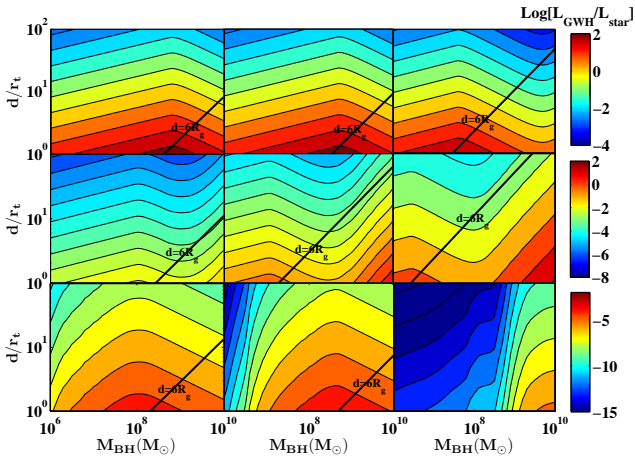}
\advance\leftskip-0.5cm
\caption{Ratio of the peak GW heating luminosity to the intrinsic stellar luminosity. The horizontal axis shows the mass of the SMBHB, and the vertical axis plots the distance (d) between the star and the SMBHB in units of the tidal radius ($r_t$). First row: model 1, 2, 3; second row: model 4, 5, 6, third row: model 7, 8, 9. Solid black line indicates where the distance between the star and black hole binary is 6 $R_{\rm g}$, the radius of the innermost stable circular orbit (ISCO) around a non-spinning black hole. In the last panel, the points in the figure lie out of 6 $R_{\rm g}$, and so the black line is not shown. The first two rows correspond to $0.1\,\Msun$ stars with metallicity $Z=0.16$ and $Z=0.01$ respectively, and the last row corresponds to $100\,\Msun$ stars. GW heating is most significant for high metallicity low mass stars.}
\label{fig:f2}
\end{figure}

Next, we discuss the heating effects in accretion disks. For $\alpha$ and $\beta$ disks, we solve Eqs.~(\ref{e:fdisk1}) and (\ref{e:fdisk2}) for the heating flux, and plot the heating light curve of the disk due to GW heating in Figure 3. The accretion disk is punctured with an inner hole. This geometry is essentially ``frozen'' during the final GW merger timescale with a gap radius $\gtrsim 100 M$ for $\alpha$--disks \citep{Milosavljevi05}. Recent MHD simulations by \citet{Noble2012} indicate that the stresses  may be enhanced in a binary, such that gap decoupling occurs further in, at $20 R_{\rm g}$. We optimistically adopt this value for our estimates, which implies a larger heating rate than that for a larger gap radius.
We integrate over the accretion disks between the inner and outer boundary. We set the latter to $2\times10^4R_{\rm g}$, but this value does not influence our result as the heating in the outer accretion disk is negligible. We include the different light travel time from different accretion disk surface elements along the line of sight.
Our calculation of the heating in the accretion disks improves the simplified treatment of \citet{Kocsis08} by including the dependence of viscosity on the ratio of the GW driving period to the largest eddy turnover time, which suppresses the dissipation of GWs. We consider two cases in this plot. Following the perturbative turbulence derivation by \citet{Goodman97}, the power-law index is 2 for Kolmogorov turbulent scaling, and $\frac{11}{9}$ according to MHD disk simulation by \citet{Flock11}. The eddy turnover time increases rapidly as the radius increases, and so the suppression of the GW heating is less significant for disks truncated closer to the SMBHB. Therefore, the heating luminosity is more significant for disks that are truncated closer to the SMBHB.

% \textbf{Because the GW luminosity is lower and the viscosity suppression is larger at larger distances from the SMBHB, the heating effect is less when the inner radius of the accretion disk is larger. Our estimation assumes that the inner radius of the accretion disk is at $20 R_{\rm g}$. This is supported by the recent MHD simulation by \citet{Noble2012}. Using the standard $\alpha$ and $\beta$ accretion disk models, the inner radius of the accretion disk is estimated to be $\sim 200 R_{\rm g}$ by \citet{Milosavljevi05} and $\sim 1000 R_{\rm g}$ by \citet{Schnittman08, Kocsis11}. The inner radius estimated by the simulation is smaller because the internal stress produced by the MHD turbulence is larger \citep{Noble2012}. In addition, the inner radius of the standard disk models may be over estimated due to the following: first of all, it is possible to have gas inside the gap due to periodic accretion \citep[e.g.][]{Chang}. Secondly, the standard disk models neglect the 3-dimensional overflow of the gas inside the disks and neglect the accumulation of gas at the inner radius of the accretion disk. These neglected effects may increase the viscosity in the inner edge of the disk and decreases the inner radius. Therefore, we set the inner radius of the accretion disk to be at $20 R_{\rm g}$.}

\begin{figure}
\includegraphics[width=3.3in, height=2.7in]{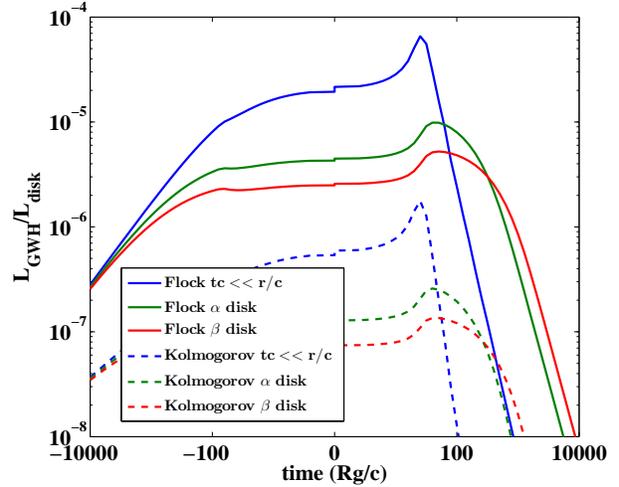}
\caption{The excess luminosity relative to the disk luminosity due to GW heating on an accretion disk (inner disk truncated at $20~R_{\rm g}$) before ($t<0$) and after ($t>0$) the binary coalescence event. The time axis is shown on a logarithmic scale at both negative and positive values (in units of $R_{\rm g}$). The SMBHB mass is $10^7M_{\odot}$. Solid lines corresponds to the frequency dependence $(\tau_{\rm GW}/2 \pi \tau_l)^ \frac{11}{9}$ derived according to the energy spectrum of accretion disk based on MHD simulations by Flock et al. (2011), and the dashed lines correspond to the scaling $(\tau_{\rm GW}/2 \pi \tau_l)^2$, assuming Kolmogorov turbulence.}
\label{fig:f3}
\end{figure}

\section{Discussion}
\label{s:conc}
In this paper, we considered the dissipation of GWs in an accretion disk or stars surrounding a SMBHB. We have found that the GW heating luminosity of the accretion disk and stars are low, and make no significant EM flare relative to their intrinsic luminosity except for low mass stars ($\sim 0.1\Msun$). The integrated excess luminosity from heated low mass stars is too low to be observed in galactic nuclei as they are faint. Assuming a Bahcall-Wolf distribution of stars or assuming a collision timescale larger than 1\,Myr, we find that only a few stars are expected to be within 5 tidal radii of a coalescing SMBHB, where the GW heating effect is significant. Therefore the overall brightening of the stellar cluster is negligible.

In order to be heated significantly by GWs, the stars need to be close to the SMBHB. One possible avenue is that stars get caught in mean motion resonances (such as Trojan resonances) and move inwards as the SMBHB merge \citep{Seto10, Schnittman10}. This is only effective for SMBHB with an unequal mass ratio $q\lesssim 10^{-2}$; the stars get ejected before the coalescence otherwise. Another possibility is for stars to get captured or form in the outer parts of accretion disks, and migrate inwards by processes analogous to planetary migration \citep{Miralda05, Karas01, Levin07}.

We assumed that GW energy is dissipated locally through turbulent viscosity. The damping of shear stress by eddy viscosity in stars was found to be consistent with observations in the context of the tidal circularization of binaries \citep{Verbunt95, Meibom05}. The underlying accretion disk model is uncertain since the disk structure is unstable to both thermal and viscous instabilities. Recently, \citet{Blaes11} found that radiation-dominated disks differ significantly from the standard disk models, where the dissipation associated with the turbulent cascade and radiative damping dissipate energy non-locally. It remains to be seen whether the GW heating effect is more prominent in alternative disk models.

%-----------------------------------------------------------------

\section*{Acknowledgments}
We thank Eliot Quataert, Sterl Phinney and Paul Groot for helpful discussions. This work was supported in part by NSF grant AST-0907890 and NASA grants NNX08AL43G and NNA09DB30A. BK acknowledges support from NASA through Einstein Postdoctoral Fellowship Award Number PF9-00063 issued by the Chandra X-ray Observatory Center, which is operated by the Smithsonian Astrophysical Observatory for and on behalf of the National Aeronautics Space Administration under contract NAS8-03060.

\bibliography{msref}

\end{document}